\documentclass[%
superscriptaddress,
amsmath,amssymb,
aps,
prl,
twocolumn,
]{revtex4-1}

\usepackage{graphicx}% Include figure files
\usepackage{dcolumn}% Align table columns on decimal point
\usepackage{bm}% bold math
\usepackage{hyperref}% add hypertext capabilities
\usepackage[capitalize]{cleveref}
\usepackage[caption=false]{subfig}
\renewcommand{\v}[1]{\ensuremath{\mathbf{#1}}} % for vectors
\newcommand{\gv}[1]{\ensuremath{\mbox{\boldmath$ #1 $}}} 
\newcommand{\diff}{\mathrm{d}}
\newcommand{\grad}[1]{\gv{\nabla} #1} % for gradient
\renewcommand{\div}[1]{\gv{\nabla} \cdot #1} % for divergence
\newcommand{\mean}[1]{\left< #1 \right>}
\newcommand{\laplacian}[1]{\grad^2 #1}

\newcommand{\pdt}[1]{\partial_t #1}

\newcommand{\Reynolds}{\mathrm{Re}}
\newcommand{\Reynoldsc}{\Reynolds_{\textrm{c}}}
\newcommand{\Reynoldscq}{\Reynolds_{\textrm{c},q}}

\begin{document}

%\preprint{APS/123-QED}

\newcommand{\NBI}{Niels Bohr Institute, University of Copenhagen, Blegdamsvej 17, DK-2100 Copenhagen, Denmark}
\newcommand{\PoreLab}{PoreLab, The Njord Center, Department of Physics, University of Oslo, P.O.\ Box 1048, NO-0316 Oslo, Norway}
\newcommand{\Njord}{The Njord Center, Department of Physics, University of Oslo, P.O.\ Box 1048, NO-0316 Oslo, Norway}

\title{Onset of turbulence in channel flows with scale-invariant roughness} % Force line breaks with \\

\author{Gaute Linga}
\email{gaute.linga@mn.uio.no}
\affiliation{\PoreLab}
\affiliation{\NBI}
%\author{Fran\c{c}ois Renard}
%\affiliation{\PGPG}
\author{Luiza Angheluta}
\affiliation{\PoreLab}
\author{Joachim Mathiesen}
\affiliation{\NBI}
\date{\today}

\begin{abstract}
Using 3D direct numerical simulations of the Navier--Stokes equations, we study the effect of wall roughness on the onset of turbulence in channel flow. The dependence of the friction factor on the Reynolds number, $\Reynolds$, is found to follow a generalized Forchheimer law, which interpolates between the laminar and inertial asymptotes. The transition between these two asymptotes occurs at a \emph{first} critical $\Reynolds$, $\Reynolds_c$, that depends nontrivially on the roughness amplitude. We identify the transition from subcritical to supercritical onset by looking at the dependence of the velocity fluctuations on $\Reynolds$ for different roughness amplitudes. We find that this \emph{second} critical $\Reynolds$ is comparable in magnitude to $\Reynolds_c$, implying that transitional flow is an integral part of flow in open fractures when $\Reynolds$ and the roughness amplitude are sufficiently high.
  
  % \begin{description}
  %   Secondary publications and information retrieval purposes.
  % \item[PACS numbers]
  %   May be entered using the \verb+\pacs{#1}+ command.
  %   You may use the \texttt{description} environment to structure your abstract;
  %   use the optional argument of the \verb+\item+ command to give the category of each item. 
  % \end{description}
\end{abstract}


%\tableofcontents

\maketitle

Since the early experiments by \citet{reynolds1883}, the onset of turbulence in wall-bounded flows has been an open problem in fluid dynamics with recent breakthroughs in our understanding of the flow between smooth walls \cite{avila2011,barkley2015,barkley2016,mukund2018}. In the smooth-wall limit, the onset of turbulence is via a subcritical transition, meaning that the laminar state is linearly stable and nonlinear perturbations are necessary in order to produce proliferation of self-sustained velocity fluctuations. These localised turbulent structures spread or decay and fill the system through spatiotemporal intermittency. In recent works, the subcritical transition in flows bounded by smooth walls is connected to directed percolation phase transition, and, in certain limits, it may even belong to the same universality class \cite{goldenfeld2006,goldenfeld2016,guttenberg2009friction,pomeau1986,hinrichsen2000,manneville2015,manneville2016}. Much less is known about the nature of the onset to turbulence in the presence of wall roughness. The classical Nikuradse measurements of the  friction factor in pipe flows with discrete wall asperities remain the main benchmark in this field \cite{nikuradse1933}. Recent work \cite{agrawal2019,hogendorn2018} has reported that the addition of a sufficient amount of particles to pipe flows may render the laminar base flow unstable and the transition to turbulence supercritical, directly passing to turbulence without spatiotemporal intermittency.

In this Letter, we present a first systematic study on the transition to turbulence in 3D flows bounded by rough walls that have a continuous and self-affine roughness. This can be considered as a prototypical, minimal model for flow in fractured materials. Albeit flow in open fractures  has been extensively studied computationally, it is mostly in the low $\Reynolds$ regime \cite{lo2012,skjetne1999,wang2016} or for steady state flows \cite{zou2017,wang2016,briggs2017}. In contrast, unsteady flow in open fractures is much less studied and understood \cite{skjetne1999, zou2017}, hence its impact on macroscopic transport properties remain elusive, particularly around the turbulent transition point. For instance in Ref.\ \cite{skjetne1999}, the authors simulated high-velocity time-independent flow in a \emph{2D} self-affine fracture joint, and found that the relationship between average forcing $f$ and mean flow $u_x$ was well described by a cubic form \cite{mei1991,lo2005}, $f \sim u_x + k u_x^3$ ($k$ being an empirical constant) at low $\Reynolds$, and the empirical Forchheimer law,
\begin{equation}
  f = a u_x + b u_x^2,
  \label{eq:forchheimer}
\end{equation}
at higher $\Reynolds$ ($a, b$ are empirical coefficients). 

Using 3D direct numerical simulations (DNS) of the Navier--Stokes equations, we determine robust scaling behaviour of friction factor (proportional to the mean force) with $\Reynolds$ (proportional to the mean flow velocity) and roughness amplitude in the time-dependent transient regime. Furthermore, by looking at the fluctuation-based Reynolds number (proportional to the mean-square fluctuations), we show that the transition to turbulence changes from being subcritical to being supercritical for sufficiently large roughness amplitude. 

\paragraph{Numerical setup:} 
As a simple idealisation of open fracture, we consider two identical self-affine surfaces that are shifted vertically along the $z$ axis by a fixed distance $d$. They form a channel which is periodic in the $x$ and $y$ in-plane directions. Thus the $\Reynolds$ number based on the flux through any perpendicular cross section is uniformly well-defined. This type of geometry is known as a fracture joint, resulting from mode I fracture, in contrast to a fault, where the surfaces would be shifted both vertically and in the $xy$ plane \cite{skjetne1999}.
The self-affine fracture surface denoted as a $z = h(x, y)$ \cite{bouchaud1997} is a random surface that is statistically invariant under the scale transformation
$ (x, y, z) \to (\lambda x, \lambda y, \lambda^H z),$
\cite{feder2013,barabasi1995}. Here, $H$ is the Hurst exponent, which we set to $H=0.8$, representative for most fractures in 3D \cite{bonamy2006,neuville2010}. 

We define the \emph{roughness amplitude} as the root-mean-square height deviation, $A = L^{-1} ({ \int_{0}^L \int_{0}^L h^2(x, y) \, \diff x \, \diff y })^{1/2}$. Due to the self-affine nature of the surface, the amplitude scales with the system size as $A \sim L^H$.
We therefore expect that the flow properties dependent on the roughness amplitude will also indirectly scale non-trivially with the system size. This has been investigated in the lubrication approximation (see e.g.\ \cite{meheust2003}), but is computationally much more challenging to do in 3D DNS. Due to the inherent computational complexity, we limit our study to a fixed size $L$. The roughness amplitudes have been chosen to be $A=0, 0.1d, 0.2d, 0.5d$, and $0.8d$, as compared to the channel width $d=1$.

We perform DNS of the incompressible Navier--Stokes equations;
$ \pdt {\v u} + \v u \cdot \grad \v u - \nu \laplacian \v u = - \grad p + \v f$, and 
$\div \v u = 0$, in a channel with self-affine walls using a finite element method and unstructured tetrahedral meshes \footnote{For simulation details, see the supplemental material.}.
Here $\v u$ is the velocity field, $\nu$ is the kinematic viscosity, and $p$ is the pressure \footnote{For convenience of notation, we have absorbed the constant density into the latter quantity.}. The flow is driven by a constant, uniform body force $\v f$, either to a laminar or a transitionally turbulent flow depending on the magnitude of $\v f$. The force $\v f$ is in the steady state (where the velocity is \emph{at}, or temporally fluctuating around, a constant value) compensated by the friction between the flow field and the rough walls. At the same time it controls the injected energy per time, $\int_\Omega \v f \cdot \v u \, \diff V$ ($V$ is volume), which is compensated by the (turbulent or laminar) dissipation rate, both at the walls and in the bulk. In all simulations, no-slip conditions are applied at the boundaries, $\v u = \v 0$ for $\v x \in \partial \Omega$. In order not to trigger any spurious long-lived turbulent modes, we start all simulations from below, i.e.~either at $\Reynolds=0$ or from a steady laminar or a transitional state below the sought $\Reynolds$.
%%%%%%%%%%%%%%%%%%%%%%%%
\begin{figure*}[t]
    \includegraphics[width=\textwidth]{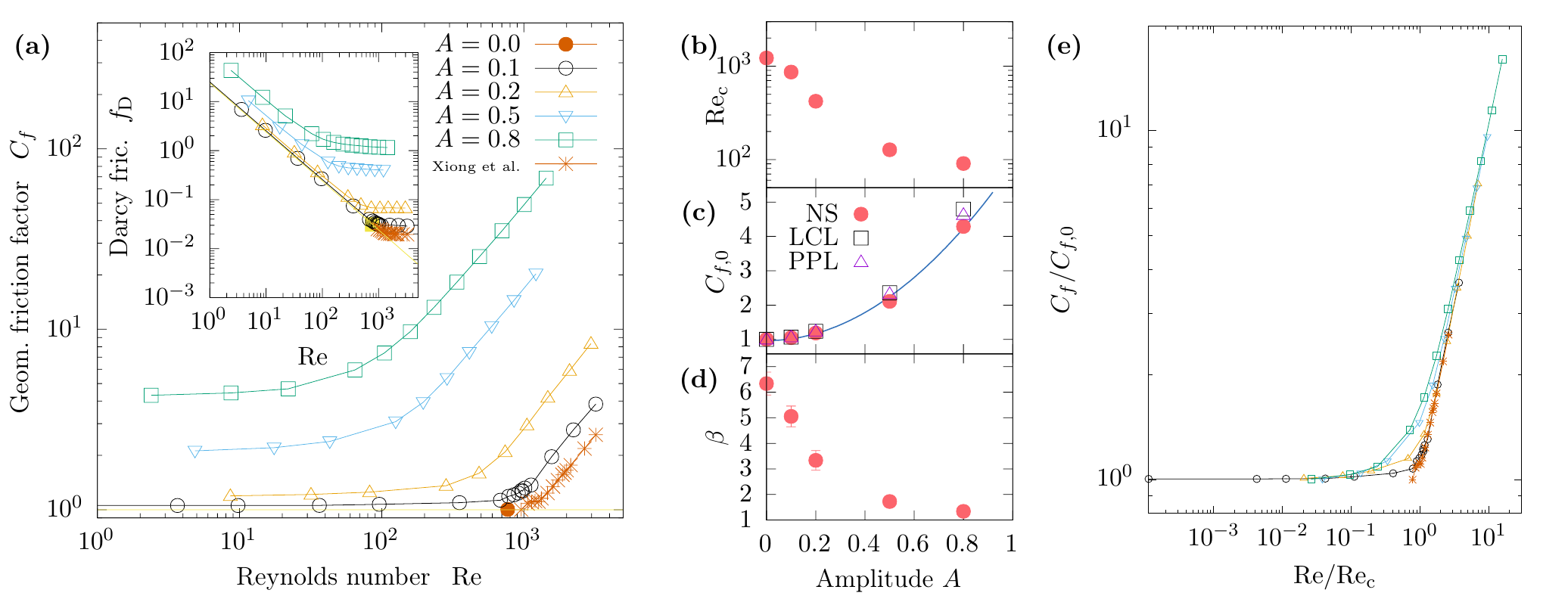}
    \caption{\label{fig:F_vs_Re}
    (a) Friction factor $C_f$ plotted against Reynolds number $\Reynolds$ for the five roughness amplitudes.
    The data for PPF marked with star symbols are taken from \citet{xiong2015}.
    Inset: Darcy friction factor $f_{\rm D}$ versus $\Reynolds$ number for the same roughness amplitudes.
    (b--d) Parameters entering into the generalized Forchheimer equation \eqref{eq:gen_forchheimer} as a function of roughness amplitude $A$.
    In (c), the numerical simulations using the Navier--Stokes equations (NS) are compared to the parallel plate law (PPL) and the local cubic law (LCL), which yields reasonable agreement. A parabolic fit to the simulation data (blue line) is shown as a guide to the eye.
    (e) Data collapse of the scaled geometric friction factor, $C_f/C_{f,0}$, as function of $\Reynolds/\Reynoldsc$ for all roughness amplitudes $A$ considered in the present work.
    In particular, we show the data presented in the main panel of (a), when $\Reynolds$ and $C_f$ are rescaled by the parameters $\Reynoldsc$ and $C_{f,0}$, respectively shown in (b) and (c).
    }
\end{figure*}

\paragraph{Friction factor:}
We define a dimensionless \emph{geometrical friction coefficient} $ C_f = {f d^2}/{(12 \nu \mean{u_x})}$ \footnote{We choose this quantity, because for Stokes flow ($\Reynolds \to 0$), this $C_f$ comes out of the equations as a \emph{purely geometric} quantity. The prefactor $1/12$ is chosen such that $C_f = 1$ for laminar flow.}
Another commonly applied quantity for pipe flows is the Darcy friction factor $f_{\rm D}$, defined through the Darcy--Weisbach relation $f_{\rm D} = {f d}/{(\tfrac{1}{2} \mean{u_x}^2)}$. These two quantities are related to each other by
\begin{equation}
  C_f = \frac{f_{\rm D} \Reynolds}{24}.
  \label{eq:Cf_def}
\end{equation}
For the special case of laminar flow between two parallel plates (PPL), we thus have $f_{\rm D} = 24/\Reynolds$.
In \cref{fig:F_vs_Re} (a), we present a diagram of the statistical steady-state relationship between $\Reynolds$ and the friction factor $C_f$ for the various roughness amplitudes. For low $\Reynolds$, $C_f$ attains a constant value dependent on the roughness $A$, whereas at  higher $\Reynolds$, there is a crossover where $C_f$ increases linearly with $\Reynolds$.
The crossover between the two regimes is where the flow becomes non-laminar and the inertial effects begin to take over. 

In order to quantify the crossover in $\Reynolds$, we consider this  functional form as a good fit to the entire range of data, at each roughness:
\begin{equation}
  \frac{C_f}{C_{f,0}} = \left[ 1 + \left(\frac{\Reynolds}{\Reynoldsc}\right)^\beta \right]^{1/\beta}.
  \label{eq:gen_forchheimer}
\end{equation}
Here, $C_{f,0}$ is the \emph{purely} geomeric friction factor, identified in the limit $\Reynolds \to 0$, while $\Reynoldsc$ is a critical Re number where the inertial effects come into play.
The exponent $\beta$ in \cref{eq:gen_forchheimer} controls the width of the transition region between the two regimes; a high exponent indicates a narrow region (fast decay) and vice versa.
Note also that when $\beta\to\infty$, $C_f/C_{f,0} = \max(1, \Reynolds/\Reynoldsc)$.
When $\beta=1$, \cref{eq:gen_forchheimer} is consistent with the Forchheimer law \eqref{eq:forchheimer}; see also \footnote{Using \cref{eq:Cf_def}, \cref{eq:gen_forchheimer} can be written in terms of the Darcy friction factor as 
\begin{equation}
  \frac{f_{\rm D}}{f_{\rm D,\infty}} = \left[ \left(\frac{\Reynoldsc}{\Reynolds}\right)^\beta + 1 \right]^{1/\beta},
  \label{eq:gen_forchheimer_2}
\end{equation}
which attains the qualitatively correct asymptotes $f_{\rm D} \sim \Reynolds^{-1}$ for $\Reynolds \ll \Reynoldsc$, and $f_{\rm D} \sim f_{\rm D,\infty} =\textrm{const.}$ for $\Reynolds \gg \Reynoldsc$, cf.\ \cite[Eq.\ (5)]{andrade1999}.
The asymptote is then given by $f_{\rm D,\infty} = {24 C_{f,0} }/{\Reynoldsc}$.}
.
Further, when $\beta = 2$, \cref{eq:gen_forchheimer} attains a quadratic correction term for $\Reynolds/\Reynoldsc \ll 1$, which is consistent with the weak inertia law.

It is thus clear that \cref{eq:gen_forchheimer} can be seen as a generalized Forchheimer equation.
While \cref{eq:gen_forchheimer} does not have a direct physical motivation, it describes the data well and provides an unbiased determination of $\Reynoldsc$ for all roughness amplitudes $A$.
$C_{f,0}$ can be read off directly from the simulation data in the $\Reynolds \simeq 0$ limit, which means that $\beta$ and $\Reynoldsc$ can be considered as the only two fitting parameters in the expression, and are readily calculated using a nonlinear least squares method.

A final test of the unified description of the data presented in \cref{fig:F_vs_Re} (a) is to inspect how well they collapse when rescaled by the parameters $C_{f,0}$, $\Reynoldsc$. In \cref{fig:F_vs_Re} (e), we plot for all simulated roughness amplitudes $A$, $C_f/C_{f,0}$ as a function of $\Reynolds/\Reynoldsc$.
For all $A$, the data is seen to follow the same asymptotic behaviour, differing only in the transition region (which in the least squares fit was captured by $\beta$).
In particular, the transition region becomes wider as the roughness is increased, consistent with the quantitative observation of the behaviour of $\beta(A)$.

%%%%%%%%%%%%%%%%%%%%%%%%%%%%%%%%%%%%%%
\paragraph{Fluctuation-based Re number:}
We separate between \emph{steady} (laminar) flow and unsteady flow, which can in principle mean both time-periodic laminar flow (where there is essentially no nonlinear transfer of energy across scales) or turbulent flow.
However, we assume that for flow over a sufficiently large rough surface (with high enough amplitude to produce detaching vortices), a time-periodic signal from a single defect will not contribute noteworthy to the overall transport properties.
Above this, there will be several (for an infinitely large domain, infinitely many) interacting `defects' that produce vortices, and thus no time-periodic signal should be found.
By using Reynolds decomposition, the velocity field $\v u (\v x, t)$ can be decomposed into its \emph{expectation value} $\overline{\v u}(\v x)$ and the \emph{velocity fluctuations} $\v u'(\v x, t)$, i.e.\ $\v u (\v x, t) = \overline{\v u} (\v x) + \v u'(\v x, t).$
Now, an indicator function for turbulent intensity can be found by defining
$ q(\v x, t) = |\v u'(\v x, t)|^2.$
Since we are primarily interested in the global presence of transitional flow, we use the space-and-time averaged indicator function $\overline{\mean{q}}$, which should only depend on $\Reynolds$ and $A$, where the error (or standard deviation) can be estimated based on the temporal fluctuations of $\mean{q}(t)$.
This leads to the definition of a fluctuation-based Re number \footnote{Note that the definition \eqref{eq:Req_def} is similar to the common definition of the Reynolds number in homogeneous isotropic turbulence.},
\begin{equation}
  \Reynolds' = \frac{\sqrt{\overline{\mean{q}}} d}{\nu},
  \label{eq:Req_def}
\end{equation}
which has the property that it is approximately zero for steady or close to steady (laminar) flow, and positive for spatially extended unsteady (transitional) flow.

\begin{figure*}[htb]
  \includegraphics[width=0.9\textwidth]{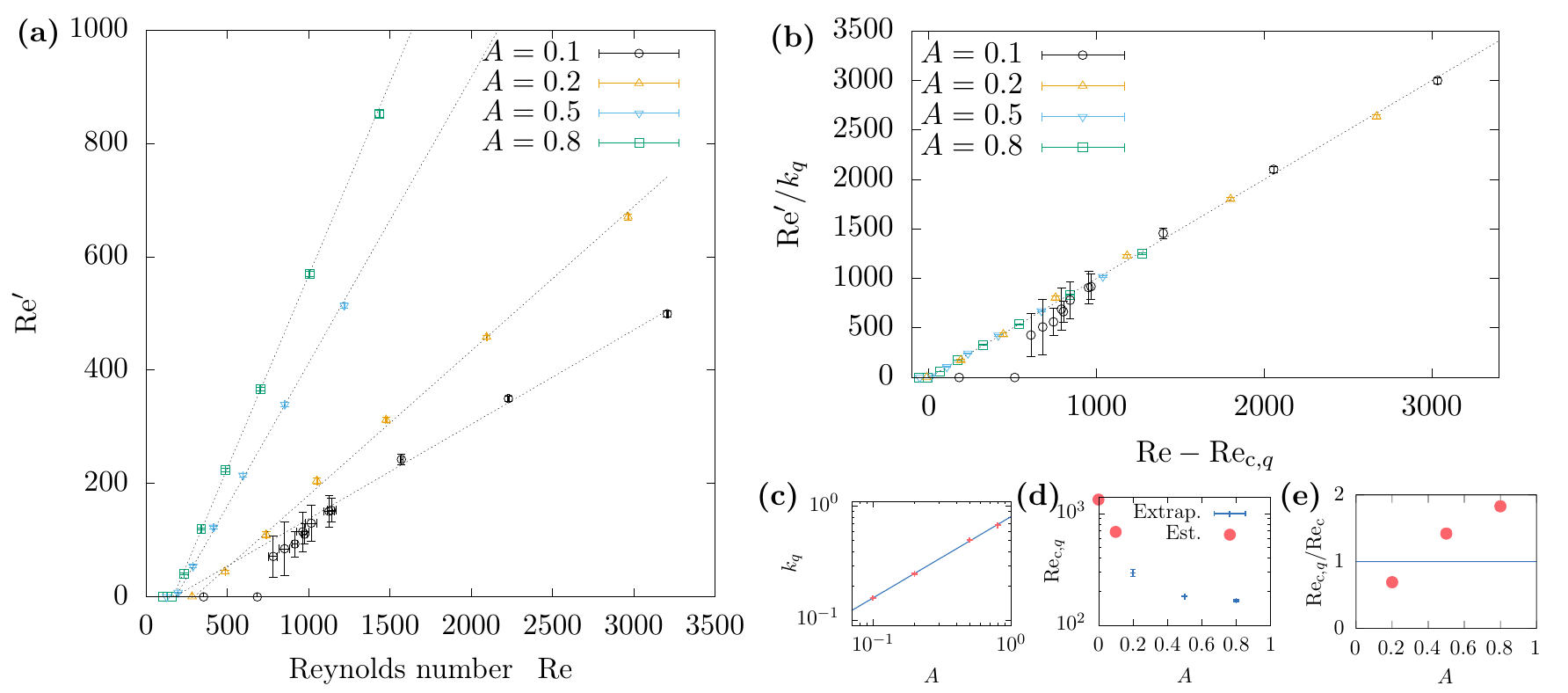}
   \caption{\label{fig:Req_vs_Re}
    (a) The fluctuation-based Reynolds number $\Reynolds'$ (indicator of turbulence) as a function of the flux-based Reynolds number $\Reynolds$, for all considered roughness amplitudes $A$. 
    (b) The same data as in (a), but scaled and shifted according to the relation \eqref{eq:Reconj}.
    (c--d) The parameters for the collapse to \eqref{eq:Reconj} determined by a least-squares fit.
    (e) The ratio between the turbulent critical Reynolds number $\Reynoldscq$ and the inertial critical Reynolds number $\Reynoldsc$.
  }
\end{figure*}

In \cref{fig:Req_vs_Re}, we show the dependence of $\Reynolds'$ on the flux-based $\Reynolds$ for all simulated roughness amplitudes $A$.
%%%%%%%%% fig%%%%%%%%%
%%%%%%%%%%%%%%%%%%%%%%%%%%%%%%%%%%
For sufficiently high $\Reynolds$, the data for all roughness amplitudes obey linear relationships.
For the lowest amplitude, $A=0.1$, the transition appears to be subcritical (as it is for $A=0$).
Around $\Reynolds \simeq 1100$, the error bars increase, indicating large temporal oscillations in the instantaneous turbulent intensity $\mean{q}$.
This indicates the presence of a metastable turbulent band which will eventually decay given sufficiently long time \cite{sano2016}.
Furthermore, the linear trend found by fitting a linear slope to the data points for which $\Reynolds' \geq \delta$ ($\delta = 10^{-3}$ is a small numerical tolerance), does not extend down to $\Reynolds' = 0$.
However, for $A > 0.1$ it does, meaning that unsteady flow is continuously produced by the boundary for $\Reynolds > \Reynoldscq$, where $\Reynoldscq$ is a \emph{second} critical Re number which quantifies the point where \emph{transitional flow} sets in, in contrast to the point of nonlinear flow resistance quantified by $\Reynoldsc$.
This suggests that the transition to turbulence changes from being subcritical to being supercritical at a roughness amplitude $A \in [0.1, 0.2]$.

Based on the adequacy of linear fits (as outlined above) to describe the $\Reynolds'(\Reynolds)$ data over roughly an order of magnitude, we propose the following relation:
\begin{equation}
  \label{eq:Reconj}
  \Reynolds' = \begin{cases}
    0 & \textrm{for} \quad \Reynolds < \Reynoldscq, \\
    k_q (\Reynolds-\Reynoldscq) & \text{for}\quad \Reynolds \geq \Reynoldscq,
  \end{cases}
\end{equation}
which should hold for amplitudes $A \gtrsim 0.2$.

\paragraph{Conclusion and discussion:}
One major advancement of this study compared to previous ones is that we focus on time-dependent transitional flow for various roughness amplitudes. In summary, the impact of a generic self-affine roughness on the macroscopic flow properties was found to be the following:
(i) The purely geometric friction factor, $C_{f,0}$ 
corresponding to the $\Reynolds \to 0$ limit of the geometric friction factor $C_f$, scales approximately quadratically with roughness amplitude $A$.
(ii) Secondly, the critical Re number $\Reynoldsc$ where inertial effects come into play decreases monotonously with $A$.
(iii) The crossover region from the constant asymptote, $C_f \sim C_{f,0}$ for $\Reynolds \ll \Reynoldsc$, to the linear asymptote
$C_f \sim \Reynolds$ for $\Reynolds \gg \Reynoldsc$, can be described by a generalised Forchheimer equation \eqref{eq:gen_forchheimer}.
The velocity fluctuations associated with transitional flow turn out to have a pronounced effect, and in particular they appear at a second well-defined $\Reynoldscq$, which has a qualitatively similar dependence on $A$ as $\Reynoldsc$, and is larger than $\Reynoldsc$ for sufficiently high $A$.
Thus, there is then a region $\Reynolds \in [\Reynoldsc,\Reynoldscq]$ where inertial effects are present but the flow remains laminar.
This implies that turbulent effects must be accounted for in simulations on larger scales already at such moderate $\Reynolds$.
Finally, our simulations and subsequent analysis suggest the turbulent transition in a rough channel goes from being subcritical (at low $A$) to being supercritical at some critical amplitude $A_{\rm c} \in [0.1,0.2]$.
This behaviour is consistent with recent observations on particle-laden pipe flows \cite{agrawal2019,hogendorn2018}, where the transition also exhibits a change from being sub- to supercritical at high particle densities. In our setup, the disordered boundary roughness plays an analogous role to particle density, rendering the laminar base flow unstable at sufficiently high $\Reynolds$.

A limitation of the present work is that we have, due to computational limitations, considered only a single realisation of a self-affine surface and varied only the roughness amplitude. In order to investigate the robustness and possible universal aspects of the present work, future research should not only consider ensemble averages of self-affine surfaces, but also of other types of roughness (e.g.\ Nikuradse-type roughness \cite{nikuradse1933,thakkar2018}).
Indeed, there is a possibility that our results are sensitive to the largest obstacle in the domain.
A second limitation, related to this, is the question of scale.
In our simulations, the domain size was fixed to $L=10 d$, while it is known that transport properties of self-affine channels scale nontrivially with the system size \cite{meheust2003}.
Future work should therefore critically reexamine whether the functional forms found here are valid regardless of $L$.
Finally, to properly quantify the universality class of the transition, significantly larger domains are needed.
As a comparison, the length scale of the domain considered in a recent study of Waleffe flow  \cite{chantry2017} was roughly equivalent to $L \simeq 1280 d$ (in our units).
Such domain sizes are out of reach with the finite element method presented herein, and an alternative route might be to follow in the lines of Ref.\ \cite{ishida2017}, who instead of resolving the complex boundary directly, used an effective body force to model boundary friction \cite{busse2012}.
However, this way of modelling roughness cannot produce vortices that are released into the bulk, and is thus invalid when the roughness amplitude is sufficiently large.

\begin{acknowledgments}
  The authors thank Anier Hernandez-Garcia and Mads H.\ A.\ Madsen for helpful discussions.
  This project has received funding from the European Union's Horizon 2020 research and innovation program through Marie Curie initial training networks under grant agreement 642976 (ITN NanoHeal), and from the Research Council of Norway through its centers of Excellence funding scheme, Project No.\ 262644.
\end{acknowledgments}

\bibliographystyle{apsrev4-1}
\bibliography{references}

\end{document}